\pdfoutput=1
\RequirePackage{ifpdf}

\documentclass[cits,11pt,a4paper]{article}
\usepackage{jinstpub}

\usepackage{graphicx}
\usepackage{natbib}

\title{Emanation and Bulk Fluorescence in Liquid Argon from Tetraphenyl Butadiene Wavelength Shifting Coatings}

\author{J. Asaadi, B.J.P Jones, A. Tripathi, I. Parmaksiz, H. Sullivan and Z.G.R. Williams}
\affiliation{Department of Physics, University of Texas at Arlington,
Arlington, Texas 76019, USA}
\emailAdd{jonathan.asaadi@uta.edu,ben.jones@uta.edu}

\abstract{We study the stability of three types of popularly employed TPB coatings under immersion in liquid argon.  TPB emanation from each coating is quantified by fluorescence assay of molecular sieve filter material after a prolonged soak time.  Two of the coatings are shown to emanate a detectable concentration of TPB into argon over a 24 hour period, which corresponds to tens of parts per billion in argon by mass. In an independent setup, the dissolved or suspended TPB is shown to produce a wavelength shifting effect in the argon bulk.  Interpretations of these results and implications for present and future liquid argon time projection chamber experiments are discussed.}


\begin{document}
\maketitle

\section{Tetraphenyl butadiene coatings in liquid argon TPC experiments}

Liquid argon time projection chambers (LArTPCs) play a central role in modern neutrino physics \cite{Acciarri:2016smi,Szelc:2016rjm,Amerio:2004ze,Antonello:2015lea} and dark matter searches \cite{Calvo:2016hve,Brunetti:2004cf,Agnes:2014bvk,Acciarri:2015uup}.  As well as being an excellent active medium for time projection chamber operation, liquid argon is also a bright scintillator, with a yield of tens of thousands of photons per MeV.  This scintillation light is emitted in the vacuum ultraviolet (VUV) range (128 nm) which presents challenges for its detection. Although argon itself is highly transparent at this wavelength, the majority of commercially available optical detectors such as SiPMs and photomultiplier tubes are not sensitive in this spectral range. 

Although some devices now exist with VUV sensitivity \cite{Erdal:2016jhl,Igarashi:2015cma,Zabrodskii:2015pda}, this problem has traditionally been solved in large-scale systems by employing a wavelength shifting coating to convert VUV light into a visible range where it can be detected by conventional sensors.  One of the most commonly used fluors is the organic compound tetraphenyl butadiene (TPB).  Various properties of TPB that are relevant to operation of LArTPCs have been studied, including its efficiency at 128~nm \cite{Benson:2017vbw,Gehman:2011xm,Francini:2013lua}, photo-degradation rates and mechanism \cite{Jones:2012hm,Chiu:2012ju,acciarri2013aging} and emission time profile \cite{Segreto:2014aia}.  However, a full understanding of the behaviour of TPB in running experiments is still evolving.  Given the reliance of future particle physics programs on the proper performance of TPB coated elements over many years, a strong understanding of the long term stability and behaviour of TPB in liquid argon is vital.  

In this work we study the stability of TPB coatings in liquid argon.  A primary goal is to establish whether TPB remains solidly affixed to coating surfaces, or rather may become dissolved or suspended in the argon bulk, for several commonly used types of coating in neutrino detectors.  Previous work has established that TPB coatings are unstable, and perhaps partially soluble, in liquid xenon \cite{sanguino2016stability}.  If this phenomenon were similarly exhibited in liquid argon, it could have significant implications for present and future neutrino and dark matter experiments.   If TPB is lifted from surfaces into the bulk, either through solvation or as particulates in suspension, it may:
\begin{itemize}
\item  Lead to wavelength shifting behavior in the bulk, or to direct excitation transfer to dissolved or suspended TPB molecules;
\item  Create a loss of performance of coatings over time;
\item  Cause TPB deposition onto surfaces other than those that are typically considered active.
\end{itemize}
These effects may lead to detrimental effects on experimental performance, or in some cases be put to constructive use if well controlled and understood.  

This paper is organized as follows.  In Sec,~\ref{sec:Coatings} we describe the coatings under study and their uses in experiments.  In Sec.~\ref{sec:Comparison} the relative stability of these coatings in liquid argon is studied, with TPB loss and subsequent accretion in filters demonstrated and quantified.  Then in Sec.~\ref{sec:Loaded}, light emission from TPB-loaded argon is measured, demonstrating that TPB steadily detaches from the surface into the bulk, and maintains its wavelength shifting character in this process.  Finally Sec.~\ref{sec:Discussion} presents a brief discussion of the interpretation and implications of our results.

\section{Coatings tested in this study \label{sec:Coatings}}

Three types of coating were used in this study, which reflect three types of TPB application widely employed in large LArTPC experiments for neutrino physics. Samples were acquired directly from the  collaborators who prepared them for their respective experiments and strongly resemble those used in operating detectors \cite{PrivateCom}.  These are:

{\bf Evaporatively coated foil (``Foils'')}.  Such foils resemble those used in the LArIAT experiment \cite{Spagliardi:2017rec,Adamson:2013/02/28tla}.  To form these coatings, a thickness of 300 $\mu g$cm$^{-2}$ TPB is evaporatively deposited onto highly reflective VIKUITI sheeting \cite{Langenkamper:2017icg} at 220$^\circ$C in vacuum. This leads to a bright white, highly efficient but somewhat mechanically fragile coating.

{\bf Over-saturated TPB in polystyrene painted coating on acrylic (``TPB+PS'') }.  This coating resembles the TPB-coated plates of the MicroBooNE experiment, prepared according to the recipe described in \cite{Acciarri:2016smi,Ignarra:2014yqa}.  To form these coatings a solution of 1~g TPB and 1~g polystyrene is prepared in 50~ml toluene with small addition of ethanol as a surfactant. As the toluene dries, the TPB becomes over-saturated in the polystrene matrix, leading to large white crystals of TPB on the plate surface.  This produces a milky white, high efficiency coating that is more mechanically robust against scratches or abrasions than the evaporative coatings.

{\bf High surface quality TPB coated acrylic (``Lightguides'')}. This coating resembles those used in the light guiding paddles \cite{Bugel:2011xg,Moss:2014ota,DenverMuons,Howard:2017dqb} of the MicroBooNE experiment \cite{Acciarri:2016smi} and proposed for the SBND \cite{Antonello:2015lea}, DUNE \cite{Acciarri:2015uup} and ProtoDUNE \cite{Abi:2017aow} experiments.  The coatings in this study were prepared similarly to those in Ref.~\cite{Moss:2016yhb}. In this case 0.1~g acrylic and 0.1~g TPB were dissolved in 50~ml toluene and a UV-transmitting acrylic bar is soaked in the solution and then drawn. This produces a surface with a high optical quality but much lower wavelength shifting efficiency than either evaporative or TPB+PS coatings.  This surface finish allows such coatings to be used in guiding light from a large surface area to a small number of optically sensitive devices.  In well-prepared coatings, the TPB layer is effectively invisible and mechanically robust.

\section{Comparison of coating robustness by fluorescence assay \label{sec:Comparison}}

\begin{figure}[t]
\begin{center}
\includegraphics[width=0.7\columnwidth]{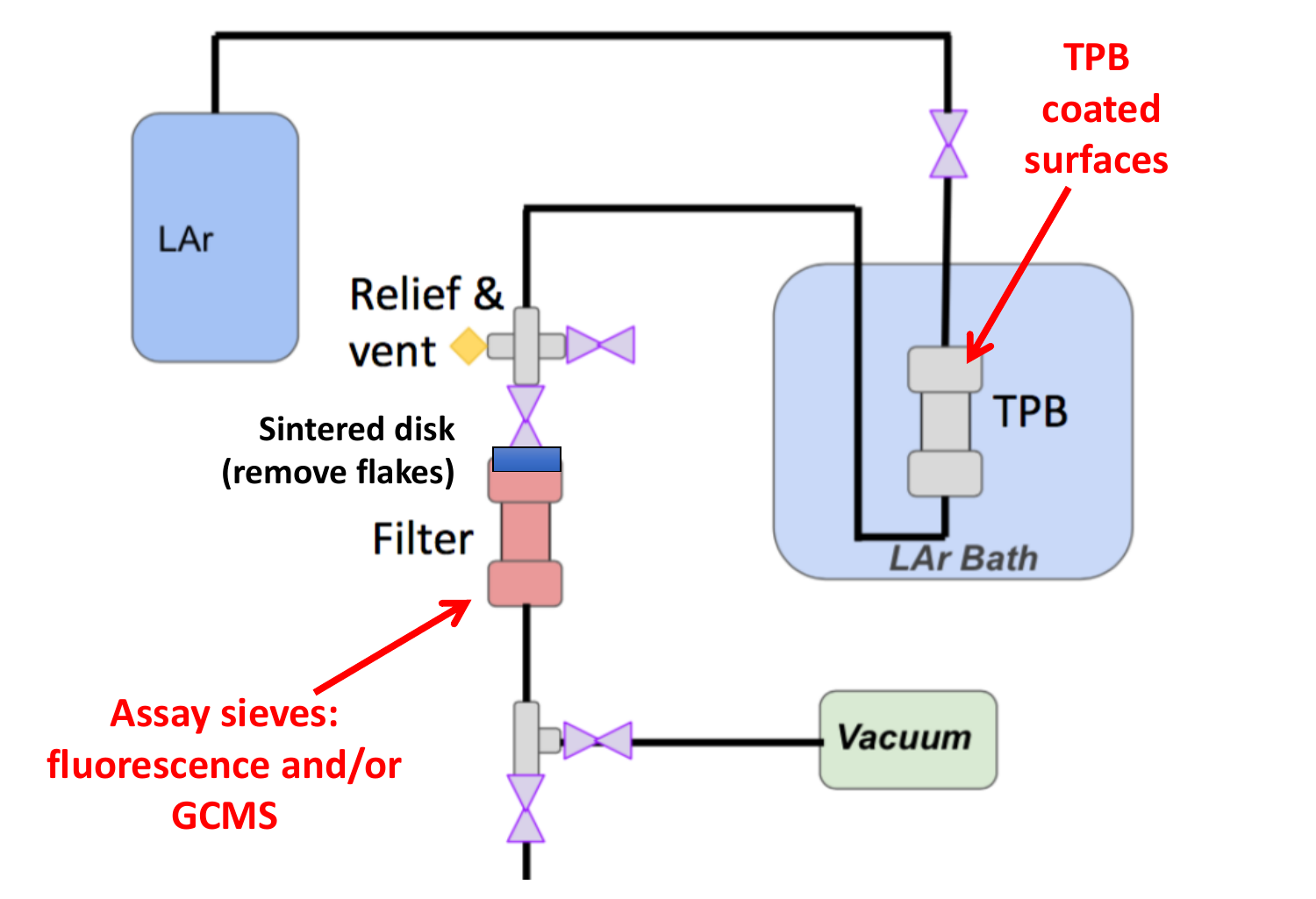}

\caption{Labelled schematic of the apparatus used to study TPB emanation and collection in the filters. \label{fig:AssayDiagram}}
\end{center}
\end{figure}

If TPB does emanate from coatings into either suspension or solution in argon, the filtration systems employed to clean this argon may remove it.  In particular, molecular sieves provide a large surface area of material onto which suspended TPB molecules or particles may adsorb.  To test the robustness of TPB coated surfaces in argon, a small cylinder of volume 119~cc containing a TPB coated element of $\sim$103 cm$^{2}$ coated surface area was filled with liquid argon (AirGas Ultra High Purity, $\leq$1 ppm O2, $\leq$1ppm H2O, $\leq$5 ppm N2) and allowed to sit for a specified number of hours.  After this time has elapsed, the liquid was driven by a back-pressure of clean argon through a column of Sigma Aldrich 4A molecular sieves between two sintered steel disks.  The argon was dumped into a dewar for evaporation to atmosphere, then the system was isolated and allowed to warm up to room temperature. Prior to each run the internal surfaces were cleaned with toluene, and the system was evacuated using a turbo-molecular pump to vacuum quality <10$^{-4}$ Torr.  Toluene will be our solvent of choice throughout this work, given its known excellent solubility for TPB.  A sketch of this system is shown in Fig.~\ref{fig:AssayDiagram}.

\begin{figure}[t]
\begin{center}
\includegraphics[width=0.65\columnwidth]{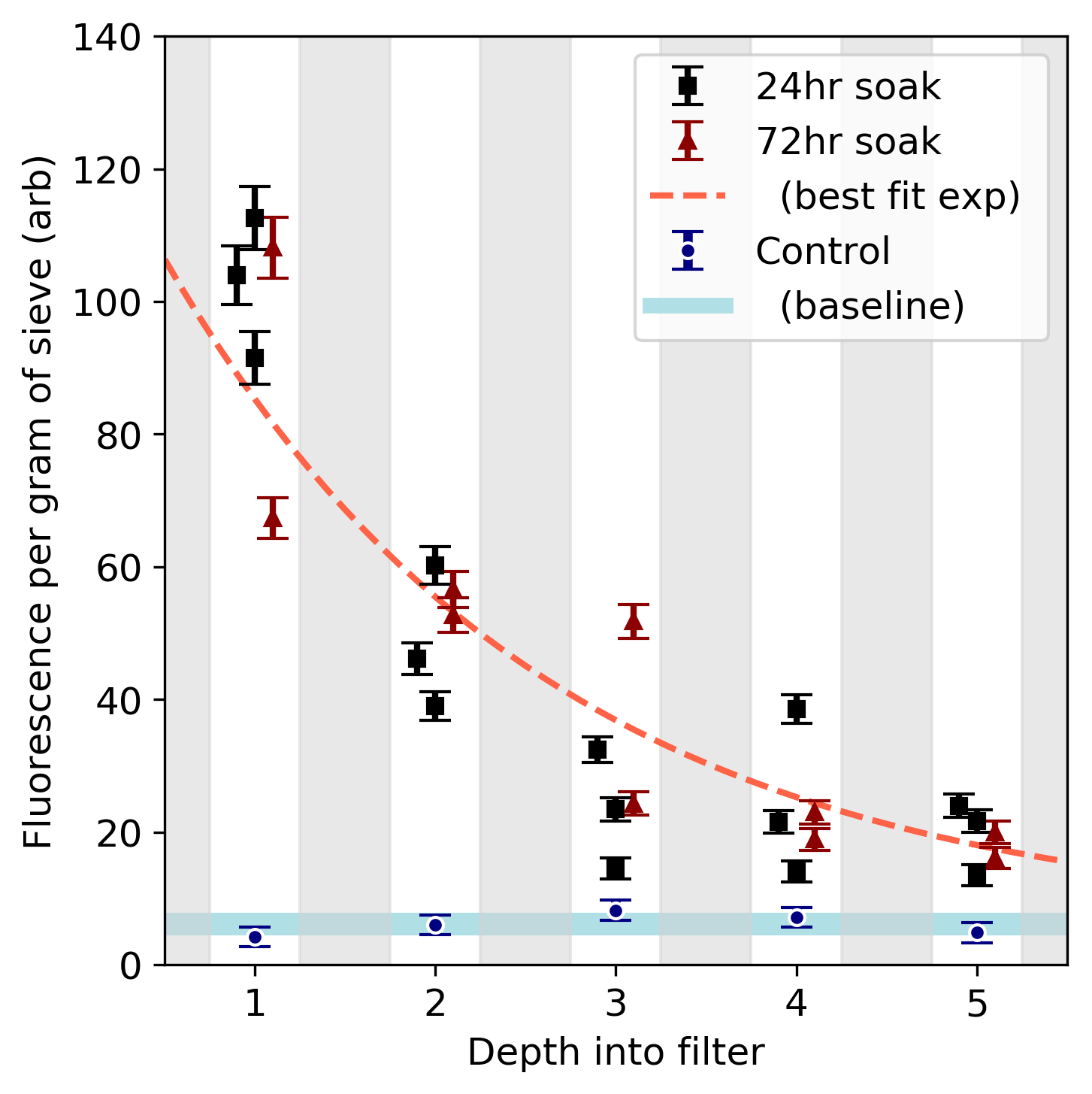}

\caption{Fluorescence intensity vs depth into the filter for the various foil samples. \label{fig:FluorescenceVsDepth}}
\end{center}
\end{figure}

Once at room temperature, the filter column was opened and the sieve material extracted in layers with approximately 90~g each, which were transposed into clean beakers. The full filter column contained approximately 6 layers, and the first five were included in this study, the latter being discarded due to its non-uniform mass depending on filter fill.  Each beaker was capped with clean aluminum foil and shaken to mix the material. A small quantity from each sample was discarded to produce samples of relatively uniform mass of 85$\pm$2~g, and each was weighed to milligram precision on an analytic balance. The samples were soaked in 100 ml of toluene for 1 hour, being stirred with a clean glass rod after 0, 30 and 50 minutes. After this soak, each solution was filtered through fine-grade filter paper to remove non-soluble sieve dust, into another set of clean beakers. These smaller beakers were left in a dark fume hood overnight such that the toluene solvent completely evaporated. After evaporation a small amount of visible white residue was observed.  This residue was present with clean sieves as well as used ones, so is not believed to be directly connected to the TPB absorption.

The residue was re-solvated in 10$\pm$0.2 ml of toluene, leading to a concentrated solution of the compounds extracted from the sieves. The samples were then pipetted into clean, capped quartz vials, which were individually scanned for fluorescence at 350~nm with 5~nm emission and excitation slit widths in an Agilent Cary Eclipse spectrophotometer.  The integrated fluorescence intensity between 400 and 600~nm was recorded for each sample.  Fig.~\ref{fig:FluorescenceVsDepth} shows the fluorescence intensity vs depth into the filter for the various samples of wavelength shifting foil, normalized to the initial sieve sample mass, for illustration.  The steadily decreasing intensity as a function of position in the filter strongly supports the interpretation that the filters are removing TPB from the argon flow.  The error bars of Fig.~\ref{fig:FluorescenceVsDepth} include quantified contributions from the stability of the spectrometer (absolute scale of 1.48 in arb. units); the initial toluene volume (100$\pm$2 ml yielding a 2\% uncertainty); the fraction of toluene actually extracted from the wet sieves (measured to be 0.73$\pm$0.02 yielding a 2.7\% uncertainty); and the volume of toluene used in resolvation of the residue (10$\pm$0.2 ml yielding a 2\% uncertainty).  The uncertainty in the tested sieve mass was also evaluated, but was negligible in all cases.

To ensure that the observed fluorescence emission was indeed from TPB, three of the solvated residue samples were tested with gas chromatography coupled to mass spectrometry (GCMS) using a similar protocol to that described in \cite{Jones:2012hm}.  A strong TPB peak was observed, with no other notable dissolved species.

\begin{figure}[t]
\begin{center}
\includegraphics[width=0.8\columnwidth]{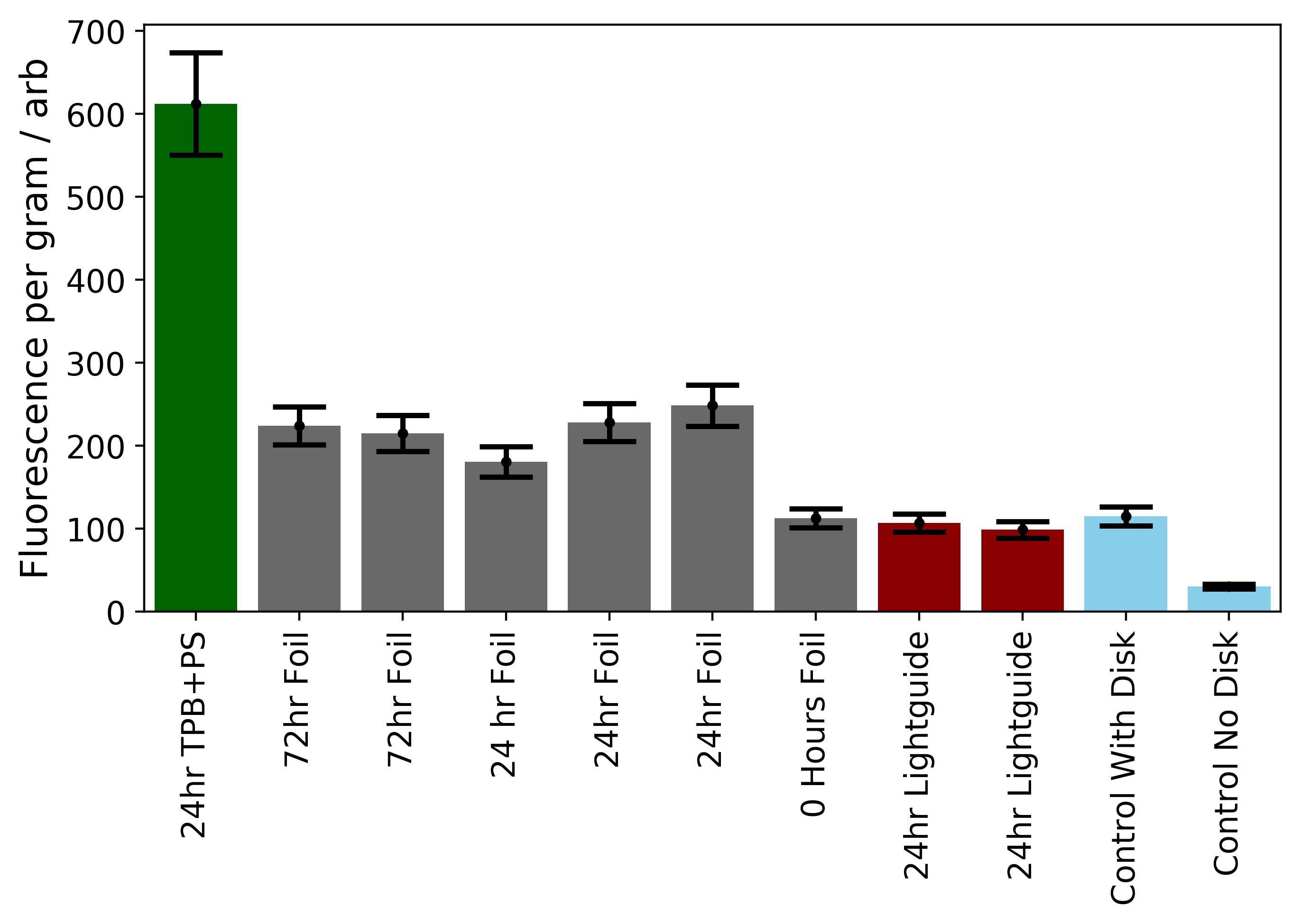}
 
\caption{Integrated fluorescence intensity for the various samples tested in this study. \label{fig:FluorescenceSamples}}
\end{center}
\end{figure}

A control run with no TPB element in front of the filter yielded an effectively fluorescence-free result, further supporting this interpretation.  Although the fluorescence from this control run was minimal (shown in Fig.~\ref{fig:FluorescenceSamples} in blue), we subsequently observed that even following a cleaning procedure involving a 30 minute soak in toluene and several washes, the sintered disks in front of the filter column retained some TPB from the previous run, which then propagated into the filter column with a clean argon flush.  This sets a lower limit on the amount of fluorescence that indicates clearly the presence of displaced TPB.  The relative intensity of fluorescence observed from each tested sample including both control configurations is shown in Fig.~\ref{fig:FluorescenceSamples}.

It is notable that neither of the light-guide samples displayed TPB in excess of the control runs with the sintered disk, and so these do not appear to emanate TPB at a level above the sensitivity of our procedure.  Also notable is that the zero-hour flush, where argon was pushed over a foil and through the column with no soak time, also does not exhibit strong fluorescence above the background level.   After either 24 and 72 hours of argon soak, the foils did cause a significant fluorescence increase in the column, though the observed increase after 72 hours was not significantly in excess of that after 24 hours.  Finally, the TPB+PS plate demonstrated by far the largest fluorescence enhancement.  Although the statistics in this test are low, the relatively stable behavior within the foil samples shows that the fluorescence  measurement  procedure is repeatable at the 10\% level. Since the foil runs were taken at various times after different prior system configurations, this uncertainty  includes the contribution from the memory effect of the sintered disk described above.  An error bar of this magnitude is  associated to each bar in Fig.~\ref{fig:FluorescenceSamples}.

\begin{figure}[t]
\begin{center}
\includegraphics[width=0.49\columnwidth]{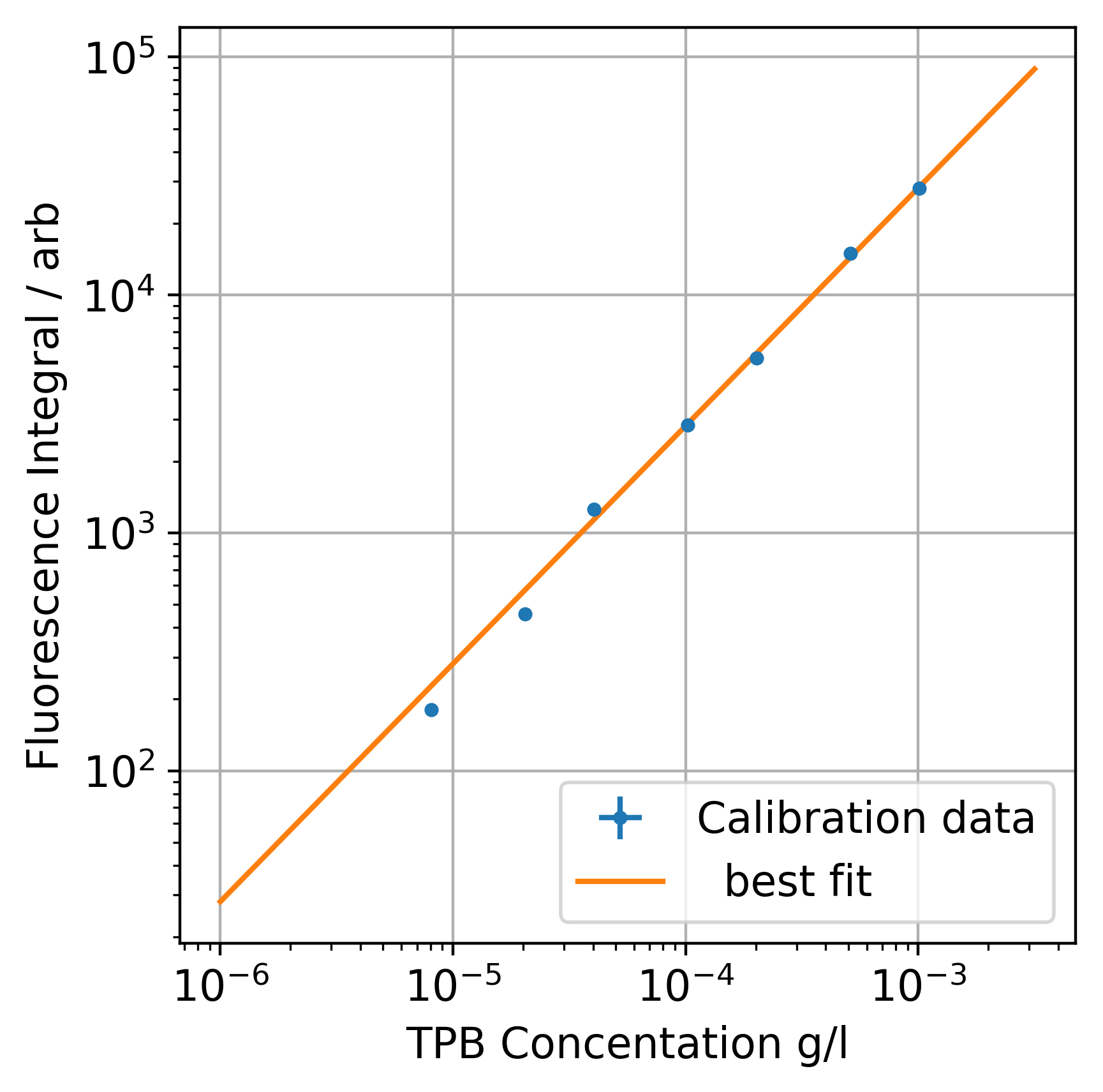}
 \includegraphics[width=0.49\columnwidth]{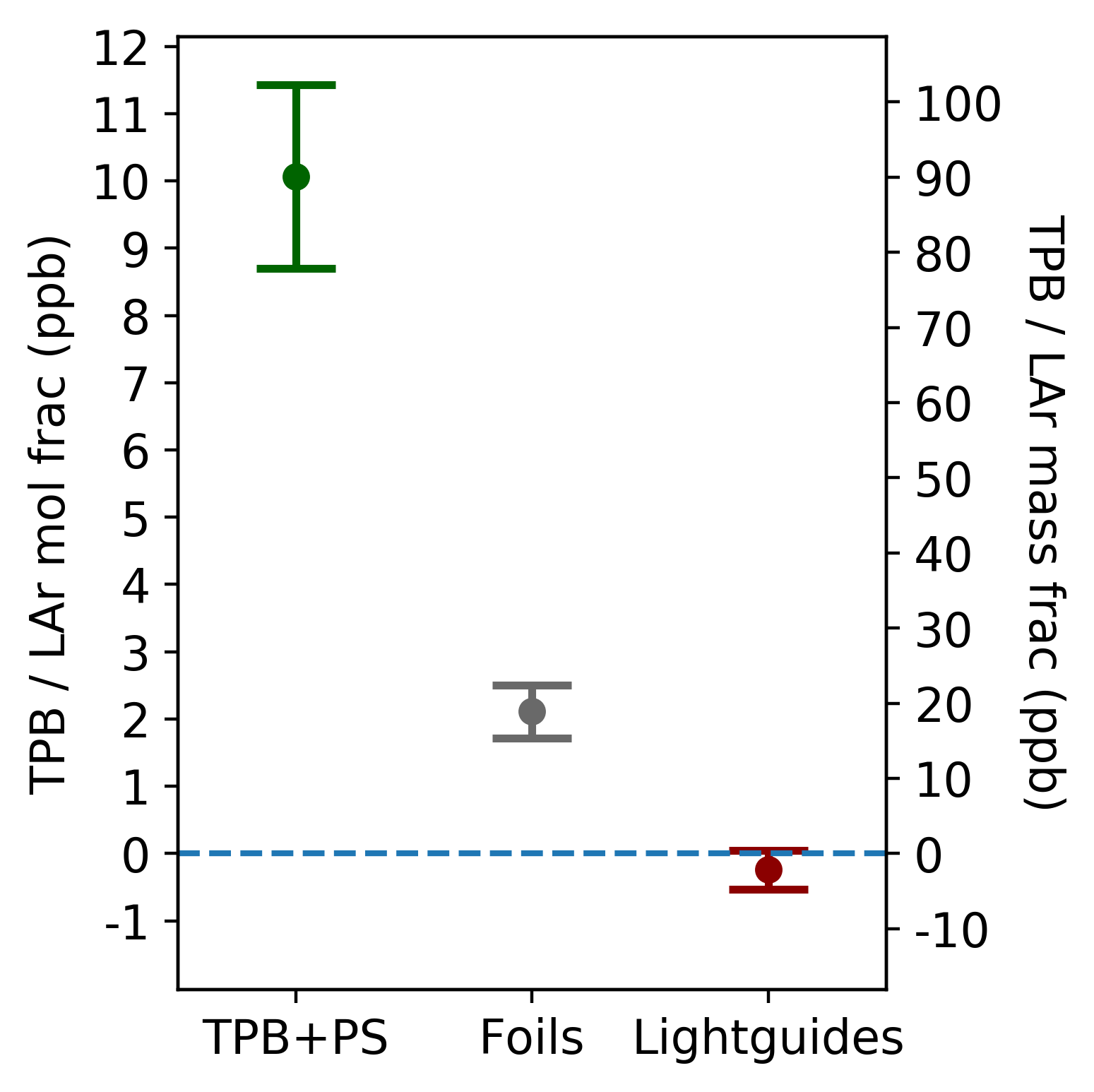}
\caption{Left: Calibration data mapping fluorescence intensity vs TPB concentration in toluene. Right: Molar and mass fractions of TPB dissolved or suspended in LAr implied by these data. \label{fig:MassFracs}}
\end{center}
\end{figure}

The actual mass of TPB absorbed into the sieves can be determined from the fluorescence measurements by comparison with calibration samples.   Samples of TPB in toluene were prepared and scanned using the same procedure as the sieve residue.   These data are shown in Fig.~\ref{fig:MassFracs}, left.  A straight line is fit to the data, allowing the fluorescence intensity of be interpreted as a measurement of the mass of TPB dissolved in toluene.  This can then be used to infer the mass density of TPB in the original liquid argon volume after the soak.  The fluorescence integral on the vertical axis of Fig.~\ref{fig:MassFracs}, left is in the same units as the y axis of Fig.~\ref{fig:FluorescenceVsDepth} after multiplying by the mass in grams ($\sim$85~g), thus the relevant range for this study is 10$^3$-10$^4$ in these units.   Fig.~\ref{fig:MassFracs}, right shows the estimated concentration of TPB in liquid argon apparently emanated by each sample type inferred from these data.  The bulk concentration may also have dependencies on parameters such as exposed surface area and coating history, so these quantities should be considered as an example rather than a universal property.  Each concentration measurement shown in Fig.~\ref{fig:MassFracs} represents only a lower limit on the TPB present in the argon, since an unknown amount will be deposited onto surfaces other than the assayed material, including the sintered disk and the vessel walls.  These data suggest that TPB emanates into argon at the tens of ppb level by mass, from both the foil and TPB+PS coatings.  The emanation from the light guide coating, if present, is below our sensitivity due to the memory effect of the sintered disk.

\section{Light emission from TPB-loaded argon \label{sec:Loaded}}
In order to test if the TPB emanated from these coatings could be a source of extraneous light in LArTPC experiments, we set up an apparatus to establish whether visible light could be detected from an ultraviolet source when TPB coated material had been allowed to soak in a pure argon bath. The experimental setup used four Hamamatsu S10362-11-050P MPPCs situated at the bottom of a cylindrical cryostat with a total volume of $\sim 70,600$ cm$^3$. Below the MPPCs, TPB coated material was enclosed between two pieces of G10 such that the wavelength shifting surface was not optically apparent to either the light source or the MPPCs. Light from a QPhotonics UVTOP 255 nm LED \cite{UVLED} was delivered into the cryostat via a UV-transmitting fiber optic. This fiber optic was aligned to shine onto a non-reflective, black surface $\sim 15$ cm above the MPPC.  This volume is filled with high purity liquid argon, and sits within an outer, open liquid argon bath for refrigeration, while the visible light yield in the internal cryostat is monitored.


Measurements were made by recording the number of MPPC signals above threshold every thirty seconds for a period of twenty minutes. For each measurement, the average of the number of counts in the thirty second interval and its standard deviation is reported. These measurements were repeated every twelve hours over a forty eight hour period for each of the configurations reported below.

\begin{figure}[t]
\begin{center}
\includegraphics[width=0.95\columnwidth]{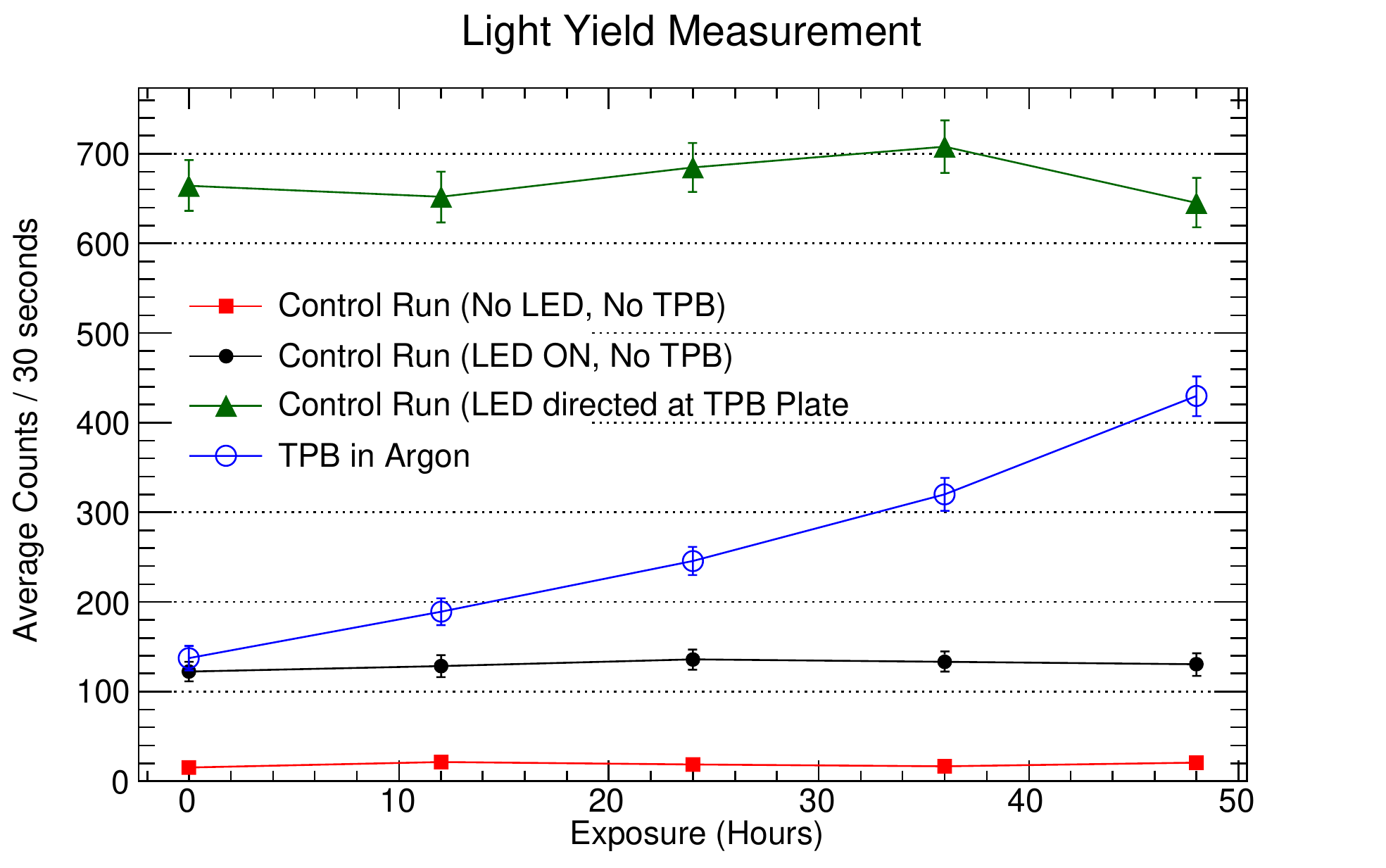}
\caption{Light yield vs time in LAr.  See text for detailed description. \label{fig:DirectLightYield}}
\end{center}
\end{figure}

At the beginning of each run, the LED was disconnected from the fiber optic to establish the rate of accidental background events. This background category includes events deriving from stray ambient light, electronics noise, MPPC dark current, or scintillation from radioactivity or cosmic rays.  The ambient rates were consistent across all measurements in the multi-day data taking period, with 18.32 counts $\pm 4.45$, shown in Fig.~\ref{fig:DirectLightYield} in red for the control run as ``No LED, No TPB''. 

A benchmark measurement was made with no TPB coated element in the system, but the LED connected and pulsed. The system was kept in this configuration for forty-eight hours to confirm the stability of the system. Some stray light was observed in this case, with a 130.0 $\pm 11.7$ counts observed (shown in Fig.~\ref{fig:DirectLightYield} in black as ``LED ON, No TPB'' ).  This light is believed to derive from the blue tail of the UV LED reflecting in the vessel, which although much less intense than the UV spectrum, is within the sensitivity range of the MPPC.  As a further benchmark, the vessel was warmed up and the non-reflective surface was replaced with TPB+PS plate $\sim 6$ cm from the fiber optic, directly illuminated. The vessel was filled with argon and measurements were again made over a forty-eight hour period with 670.8 $\pm 28.2$ counts was recorded.  This is shown in green on Fig.~\ref{fig:DirectLightYield} as ``LED directed at TPB plate'' and provides a reference point corresponding to a bright, efficient wavelength shift.  Subsequent measurements can then be referenced against these two extreme cases.

After these initial characterizations, the vessel was cleaned and samples of TPB+PS with equal surface area to those used in Section \ref{sec:Comparison} were inserted into the chamber below the MPPC's. The aim of this trial was to search for evidence for scintillation light due to TPB in solution/suspension. The vessel was again filled with argon with the fiber aligned to the black, non-reflective beam stop. A measurement of the MPPC rate was made immediately after filling, for LED on and off. The LED on rate was found to be 137.0 $\pm 13.8$, consistent with the control ``NO TPB'' measurement and the LED off rate was found to be 17.2 $\pm 4.3$, also consistent with the control measurement. These measurements were repeated every twelve hours for forty-eight hours for LED on and off.  The LED ON/OFF data are shown with blue circles and red squares respectively, in Fig.~\ref{fig:DirectLightYield}. The amount of light observed for a fixed LED intensity is observed to grow gradually over the 48 hour time window, consistent with the hypothesis of TPB emanation from the coatings into the bulk, as observed in the filter assay studies. To test the stability of the system, the vessel was evacuated, the TPB samples removed, and the setup cleaned before being filled with fresh argon for a further benchmark set of LED on and LED off measurements. The LED on rate was found to be 131.3 $\pm 11.2$ and the LED off rate was 16.4 $\pm 4.7$, both of which are consistent with the previous control measurements, strongly suggesting that the additional light originates from TPB emanation.

It is notable that the timescale of emanation is multiple hours, consistent with the null results from the ``zero hour'' filter assay and non-null results from the ``24 hour'' assay test in Sec.~\ref{sec:Comparison};  and also that the displaced TPB appears to be not only a visible light source but also a rather bright one with 429.2 $\pm 22.3$ counts after 48 hours, around half the intensity of the directly illuminated coating.

Preliminary studies of pulse timing indicate components of the wavelength shifted light with long time constants, in the milliseconds range.  This may represent a phosphorescent behaviour that is suppressed when TPB is densely packed on a surface but uninhibited in weak solutions.  The details of these time constants will be studied in future work.

\section{Discussion \label{sec:Discussion}}
We have conclusively demonstrated that TPB emanates from certain coating types that are commonly used in liquid argon particle physics experiments, when submerged in liquid argon over hour to day timescales. This emanated TPB is removed from the argon bulk by molecular sieves, and detaches most strongly from coatings where the fluor is not protected by a polymer matrix.  This has possible implications for running and proposed particle physics experiments that use either painted or evaporative coatings.  

The amount of TPB that becomes dissolved or suspended is in the range of tens of parts per billion by mass.  This is sufficient to cause bulk fluorescence in the argon, and this fluorescence has been demonstrated using 255 nm excitation by LED.  The bulk fluorescence for an argon sample near saturation is found to be of similar magnitude to the brightness of an exposed coating, suggesting that this may be a significant visible light source in experiments where TPB is present.  

Although our data demonstrate TPB emanation, they are unable to distinguish between a true solvation effect or a suspension of larger particulates.  These two possibilities can be considered as points in a continuum, with colloidal suspensions of very small particles behaving almost indistinguishably from a true solution in all practical cases of interest. The previous reported instability of TPB films in liquid xenon \cite{sanguino2016stability} was also not conclusively characterized.  If that effect were to be attributed to solubility, then the expected scaling of Van-Der-Waals (VDW) forces between these two noble elements according to the London Dispersion Equation \cite{hirschfelder1954molecular} would suggest similar qualitative effects should be expected in argon, the ratio of VDW forces being around a factor of $\sim$2.6, assuming the ionization energy of TPB to be around 5~eV and the well known polarizibilities of argon and xenon atoms.

Because solubility / emanation is likely to depend on parameters including liquid flow rate, history of coating preparation and handling, fill procedure, time since installation, filtration method, and potentially also purity, evacuation, purge, or bake-out procedure, it is not possible to make quantitative predictions for any given experiment, given our present level of understanding.  However, the previously unreported phenomena presented in this paper may have significant implications in running and planned detectors, and further consideration of these effects in more targeted experimental configurations appears well motivated.

There is also a notable possibility that dissolved or suspended TPB in argon may represent an opportunity rather than a burden.  If a suitable filtration and circulation system could be implemented to maintain a specific steady-state concentration, this would remove the need for additional wavelength shifting elements such as light guides, coated plates or foils.  Such a detector with an inherently wavelength shifting noble medium could achieve higher efficiency of light collection via reduced solid-angle losses, and also enjoy tuneable optical parameters after installation, among other benefits.  Such a light detection concept may be promising for future large LArTPCs and cryogenic scintillation experiments.

\section*{Note Added to arXiv version:}
V1 of this paper as submitted to arXiv includes information on time constants of the measured light from the argon bulk. We have no reason to doubt the correctness of these measurements.  However, some checks were requested as part of the journal review process that we were not able to undertake on a short timescale.  We thus opted to remove this section from V2, in order to enable publication of the main body of the work within a suitable timeframe.  The long time constants will be the subject of more detailed investigations and reporting in future work. 

\section*{Acknowledgements}
We thank Janet Conrad and Andrzej Szelc for providing samples and useful feedback and encouragement during this work, and Flavio Cavanna for insightful comments. The equipment used in these studies was procured for the MicroBooNE, DUNE and NEXT programs.  The UTA group is supported by the Department of Energy under contract numbers DE-SC0011686 and DE-SC0017721.

\bibliographystyle{JHEP}
\bibliography{main}

\end{document}